\documentstyle[11pt,paspconf,epsf]{article}

\begin{document}

\def\vmax{$V_{\rm max}$}
\def\mueff{$<$$\mu$$>_{\rm e}$}
\def\reff{$r_{\rm e}$}

\title{The Space Density of Spiral Galaxies as function of their Luminosity,
Surface Brightness and Scalesize}
\author{Roelof S. de Jong\altaffilmark{1}}
\affil{Univ.\ of Durham, Dept.\ of Physics, South Road, Durham
DH1 3LE, UK}
\author{Cedric Lacey}
\affil{TAC, Juliane Maries Vej 30, DK-2100
Copenhagen O, Denmark}

\altaffiltext{1}{Hubble Fellow, Steward Observatory, 933 N. Cherry Ave.,
AZ 85716, USA} 

\begin{abstract} 
 The local space density of galaxies as a function of their basic
structural parameters --like luminosity, surface brightness and
scalesize-- is still poorly known.  Our poor knowledge is mainly the
result of strong selection biases against low surface brightness {\em
and} small scalesize galaxies in any optically selected sample.  We show
that in order to correct for selection biases one has to obtain accurate
surface photometry {\em and} distance estimates for a large ($\ga$1000)
sample of galaxies.  We derive bivariate space density distributions in
the (scalesize, surface brightness)-plane and the (luminosity,
scalesize)-plane for a sample of $\sim$1000 local Sb-Sdm spiral galaxies. 
We present a parameterization of these bivariate distributions, based on
a Schechter type luminosity function and a log-normal scalesize
distribution at a given luminosity.  We show how surface brightness
limits and (1+$z$)$^4$ cosmological redshift dimming can influence
interpretation of luminosity function determinations and deep galaxy counts%
%, in particular for the Hubble Deep Field
. 
 \end{abstract}


%---------------------------------------------------------------------------
\section{Introduction}

Knowing the space density of galaxies as function of their structural
parameters (luminosity, surface brightness (SB) and scalesize) is
important when:

 1) making comparisons between different galaxy samples, because
selection functions of extended resolved objects depend on at least two
structural parameters.  This becomes particularly relevant when comparing
samples at different redshifts, where (1+$z$)$^4$ redshift dimming can
give rise to strong SB biases. 

 2) testing galaxy formation and evolution models, as any successful galaxy
formation theory will have to be able to explain the spread in structural
parameters and their relative frequency in the local galaxy population.

Many papers have been devoted to the determination of the space density
of galaxies as function of their luminosity, i.e.\ the galaxy luminosity
function (for a recent review see Ellis 1997).  In many of these papers
one has conveniently ignored the possibility of strong SB biases. 
Determinations of scalesize distributions have been scarce (some notable
exceptions van der Kruit 1987, Hudson \& Lynden-Bell 1991, Sodr\'e \&
Lahav 1993, de Jong 1995) and often diameter distributions are
calculated.  Diameters distributions are not very useful in sample
comparisons, as diameters have to be measured at a certain SB level,
which might differ from sample to sample.  Realize for instance, that
there may be many galaxies that do not have a $D_{25}$, because there SB
is below 25\,$B$-mag\,arcsec$^{-2}$. 

Since the classical paper of Freeman (1970), many papers have been
devoted to the distribution of SB of disks in spiral galaxies (for
review see Impey \& Bothun 1997).  Freeman found that 28 galaxies in his
incomplete sample of 36 had disk central SB values of
21.65$\pm$0.3\,$B$-mag\,arcsec$^{-2}$ (see his review in these
proceedings).  Disney (1976) showed that the limited range in disk
central SB values might be the result of selection biases. 
 %At the same
%luminosity, high SB galaxies become to small to be selected, low SB
%galaxies disappear in the sky noise.  
 Since then several authors have argued that there seems to be indeed an
upper limit in the SB distribution near Freeman's value, but that the
distribution stays nearly flat when going to lower SB (e.g.\ McGaugh et
al.~1995; de Jong 1995).  Recently this picture has been challenged by
Tully \& Verheijen (1997), who argued that the SB distribution is
bimodal, based on $K$-band data of $\sim$60 galaxies in the Ursa Major
cluster. 

In this paper we show that one should not try to separate the
distributions of luminosity, SB and scalesize, but combine two of these
to make bivariate distributions, as any sample will have selection
biases in at least two structural parameters.

%---------------------------------------------------------------------------
\section{Correcting for selection bias}

Many methods have been devised to correct observed frequencies of object
properties for selection bias in order to obtain true space density
distributions.  We will here concentrate on the \vmax\ method, where
each object gets a weight proportional to the inverse of its maximum
sample inclusion volume (Felten 1976).  This metod is only correct if
the objects are distributed homogeneously in space, and therefore the
smallest objects in the sample should be visible at distances greater
than the largest large scale structures in the universe.  Homogeneity
and completeness can be checked with the V/\vmax\ method (e.g.\ van der
Kruit 1987 and references therein).  Accurate \vmax\ values can be
derived for each object for the modern surveys with automatic detection
algorithms on digitized data.  Each object should be artificially blue-
or redshifted and be Monte Carlo replaced at many positions in the
original data set.  The recovery fraction of the automatic detection
routine supplies the volume searched at each redshift shell and provides
information for confusion limits and Malmquist bias at the survey
limits.  For samples not selected by an automated routine from digitized
data (e.g.\ eye-selected from photographic plates), we just have to
assume that the selection criteria are well behaved when we imagine moving a
galaxy in distance. 

Moving more specifically to the distribution of structural parameters of
spiral galaxies, we will use the case of perfect exponential disks in a
diameter limited sample. More generalised descriptions can be found in
Disney \& Phillipps (1983) and McGaugh et al.~(1995). For an exponential
disk with physical scalelength $h$ and central SB $\mu_0$ we find
for the maximum distance at which a galaxy can lie before dropping out of the
sample
 \begin{equation}
d_{\rm max}\propto (\mu_{\rm lim}-\mu_0)\,h/\theta_{\rm lim},
 \label{dmax}
 \end{equation}
 with $\theta_{\rm lim}$ the sample angular diameter limit measured at
SB limit $\mu_{\rm lim}$.  As the volume where a galaxy is visible goes
as $d_{\rm max}^3$, this shows the strong selection bias against small
scalesize and low SB galaxies.  The scalesizes of spiral galaxies vary
easily by a factor of 10 (de Jong 1996).  Therefore, if all scalesizes
were equally abundant at a given SB, we would have a 1000 times more of
the largest scalesize galaxies than the smallest scalesize galaxies in a
diameter limited sample.  Luckily nature has not been that cruel to us
and there are many more small galaxies then large ones.  In the case of
the SB distribution we have not been so lucky, as the SB distribution stays
rather constant --at a given scalelength-- going to lower SB values. 
Equation\,(\ref{dmax}) shows that, at fixed scalelength, the visible
volume of a galaxy 1\,mag above the SB limit is 125 times smaller than
that of a galaxy 5\,mag above the SB limit.  In order to have some
number statistics close to the selection limit, we had better observe
hundreds of galaxies to determine a SB distribution.  Because we do
not {\it a
priori} know whether the scalesize and SB distributions are uncorrelated,
we had better make sure that we determine the SB distribution at different
scalesizes, and so we need at least  1000 galaxies. 

SB measurements are distance independent (at least on local scales); a
property that sometimes has been used to argue that one can determine SB
distributions without knowing distances.  If the distribution of $h$ is
the same at each SB level, the $h/\theta_{\rm lim}$ factor in
Eq.\,(\ref{dmax}) cancels out on average and one can make relative
volume corrections without having to know physical scalesizes/distances. 
Likewise, using total magnitude of an exponential disk
$M\propto\mu_0-5\log(h)$, Eq.\,(\ref{dmax}) can be rewritten as
 \begin{equation}
d_{\rm max}\propto (\mu_{\rm lim}-\mu_0)\,10^{-0.2(M-\mu_0)}.
 \label{dmaxM}
 \end{equation}
 Again, assuming the SB distribution is the same for each luminosity,
one can make relative volume corrections (very different from
Eq.\,(\ref{dmax})!) to calculate a SB distribution without knowing
distances.  There is no reason for the SB distribution to be independent
of either scalesize or luminosity (and we will show this is indeed not
the case) and therefore Eq.\,(\ref{dmax})\,\&\,(\ref{dmaxM}) show that
we need to know the distribution of at least one other {\em distance
dependent} structural parameter to determine the SB distribution of
galaxies.  The reverse is also true: to measure the distribution of
scalesizes or luminosities we also need to determine the distribution of
one of the other structural parameters.  In order to do so we will need
surface photometry and distances for a sample of at least $\sim$1000
galaxies.

In this paper we will use the effective radius (\reff, the radius enclosing
half of the total light of the galaxy) and the average effective SB
within this radius (\mueff) instead of the more conventional parameters
for disks, scalelength and central SB.  Using the effective parameters
has the virtue that one does not have to make assumptions about the
light distribution in the galaxy (all galaxies have an \reff, even
irregular ones) and avoids complicated bulge/disk decomposition issues. 
The distributions presented here have also been calculated for disk
parameters alone with very similar results, because most of the objects
are of late spiral type with insignificant bulge contributions.

%---------------------------------------------------------------------------
\section{Local space density distributions}

As described in the previous section, one needs accurate surface
photometry and distance estimates for a sample of at least 1000
galaxies to create bivariate distributions.  The galaxies should be
selected in a well defined, reproducible and complete way.  Data sets
obeying all of these criteria are not available at the moment, but
fortunately peculiar motion studies have produced large data sets with
accurate photometry and redshifts.  We have used the Mathewson, Ford \&
Buchorn~(1992, 1996, MFB hereafter) data set, which was selected from
the ESO-Uppsala catalog, a catalog with galaxies selected and classified
by eye from photographic plates.  We reselected a sample close to the 
MFB criteria from the
ESO-Uppsala catalog, allowing us to evaluate incompleteness in the MFB
sample (some galaxies were not observed due to foreground stars or
inability to obtain a velocity width).  We selected all galaxies from
the ESO-Uppsala catalog with type 3$\le$T$\le$8 (Sb-Sdm), angular
diameter 1.7\arcmin$\le$$\theta_{\rm maj}$$\le$5\arcmin, axis ratio
0.174$\le$b/a$\le$0.776 and galactic latitude $|b|$$>$$11\deg$.  This
resulted in a sample of 1007 galaxies, of which about 850 have $I$-band
surface photometry and redshifts. 

The luminosity, \reff\ and \mueff\ values of the galaxies were derived
from the luminosity profiles and corrected for Galactic foreground
extinction using the prescription of Schlegel et al.~(1998).  Corrections
for inclination and internal extinction were performed following a
method similar to Byun~(1992).  Distance estimates were obtained from
the Mark III catalog (Willick et al.~1997) if available, otherwise
computed from the Hubble distance, with
$H_0=65$\,km\,s$^{-1}$\,Mpc$^{-1}$. 

\begin{figure}[t]
\begin{center}
\mbox{
 \epsfxsize=9cm
 \epsfbox[100 113 490 380]{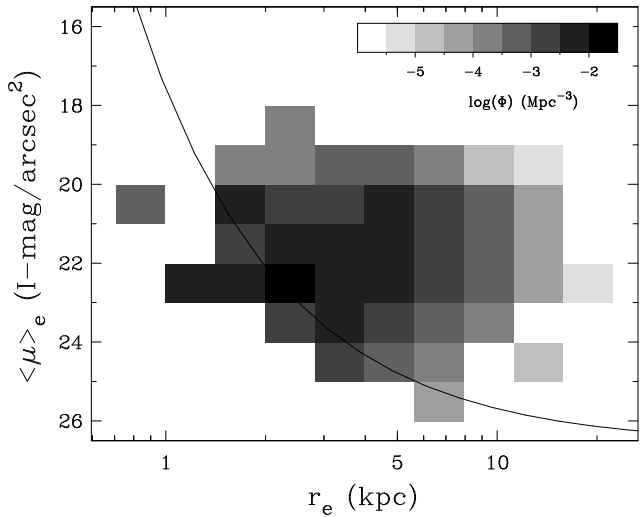}
}
\end{center}
 \caption{
 The space density distribution of Sb-Sdm galaxies as function of
effective radius and average SB within that radius.  Galaxies
with exponential disk, having structural properties indicated by the
line, can be seen out to 20\,Mpc before dropping out of the sample. 
}
 \label{bidisremuave}
 \end{figure}

Using the \vmax\ method described in the previous section, we have
calculated the bivariate density distribution in the
(\reff,\mueff)-plane, which is presented on a logarithmic scale in
Fig.\,\ref{bidisremuave}.  The paucity of galaxies in the top-right
corner of the diagram is real, large, high SB galaxies are readily
visible.  To the bottom-left of the indicated 20\,Mpc visibility line we
are hit by low number statistics; for such small, low SB galaxies we are
sampling too small a volume to have reliable statistics.  The
distribution shows a dramatic increase in galaxy space density going to
smaller scalesizes.  At a given scalesize, the SB shows a broad
distribution, peaking at about \mueff=21.5\,$I$-mag\,arcsec$^{-2}$. 
There is some indication that the peak in the distribution shifts to
lower SB at smaller scalesizes. 

%More than space density, the luminosity density of galaxies as function
%of their structural parameters tells us better in what kind of galaxy
%most of the mass in the universe is locked up.  By weighing each galaxy
%its luminosity, we can convert Fig.\,\ref{bidisremuave} into a
%luminosity density distribution, which is shown in
%Fig.\,\ref{bidisremuavelum}.  Most of the luminosity in the universe is
%provided by rather large, high SB galaxies, even though they are not the
%most numerous systems.  It is curious, but not completely surprising,
%that we happen to live a galaxy with structural properties very near the
%peak in the luminosity density distribution. 

% \begin{figure}
%\begin{center}
%\mbox{
% \epsfxsize=10cm
% \epsfbox[100 110 490 390]{premuavewI.cps}
%}
%\end{center}
% \caption{
% The bivariate luminosity density of Sb-Sdm galaxies as function
%of \reff\ and \mueff.
%}
% \label{bidisremuavelum}
% \end{figure}

%---------------------------------------------------------------------------
\section{Parametrization of the distributions}

In this section we will define a parametrization of the bivariate
distributions, as an aid to compare distributions derived from
differently selected samples or to study redshift evolution.  
%The
%parametrization can also be used in modeling where both luminosity
%and size are required (e.g.\ cross-section modeling of Lyman forest host
%galaxies). 
%
 We will follow the most simple form of the Fall \& Efstathiou (1980)
disk galaxy formation theory to derive such a parametrization (for
extended versions of the theory see e.g.\ van der Kruit 1987; Dalcanton
et al.~1997; Mo et al.~1998; van den Bosch 1998).  Galaxies form in this
theory in hierarchically merging Dark Matter (DM) halos, giving rise to
a distribution of DM halo masses described by the Press \& Schechter
(1974) theory, which formed the inspiration for the Schechter (1976)
luminosity function (LF).  We will use a Schechter LF to describe the
luminosity dimension of our distribution function. 

In the Fall \& Efstathiou (1980) model, the scalesize of a galaxy is
determined by its angular momentum, which is acquired by tidal toques
from neighbouring DM halos in the expanding universe.  The total
angular momentum of the system is usually expressed in terms of the 
dimensionaless spin parameter (Peebles 1969)
 \begin{equation} 
\lambda = J |E|^{1/2} M_{\rm tot}^{-5/2} G^{-1},
 \label{spin}
 \end{equation} 
 with J the total angular momentum, $E$ the total energy and $M_{\rm
tot}$ the total mass of the system, all of which are dominated by the
DM halo.  N-body simulations (e.g.\ Warren et
al.~1992) show that the distribution of $\lambda$ values acquired from
tidal torques in an expanding universe can be well be approximated by a
log-normal distribution with a dispersion $\sigma_{\lambda}\sim 0.5$
in $\ln(\lambda$).

A few simplifying approximations allow us to relate each of the factors
in Eq.\,(\ref{spin}) to our observed bivariate distribution parameters.  A
perfect exponential disk of effective size \reff, mass $M_d$, rotating with a
flat rotation curve of velocity $V_c$ has $J_d\propto M_d r_{\rm e} V_c$.
We assume that the specific angular momentum of the disk is equal to
the specific angular momentum of the dark halo.  From the virial
theorem we get $E \propto V_c^2
M_{\rm tot}$.  If we assume that light traces disk mass
($M_d\propto L)$ and that disk mass is proportional to total mass
($M_{\rm tot}\propto M_d$), we only need the Tully \& Fisher (1977)
relation ($L\propto V_c^\beta$, with $\beta\sim 3$ in the $I$-passband)
to link the spin parameter $\lambda$ to our observed bivariate
distribution parameters. 

 \begin{figure}[t]
\begin{center}
\mbox{
 \epsfxsize=9cm
 \epsfbox[20 170 546 670]{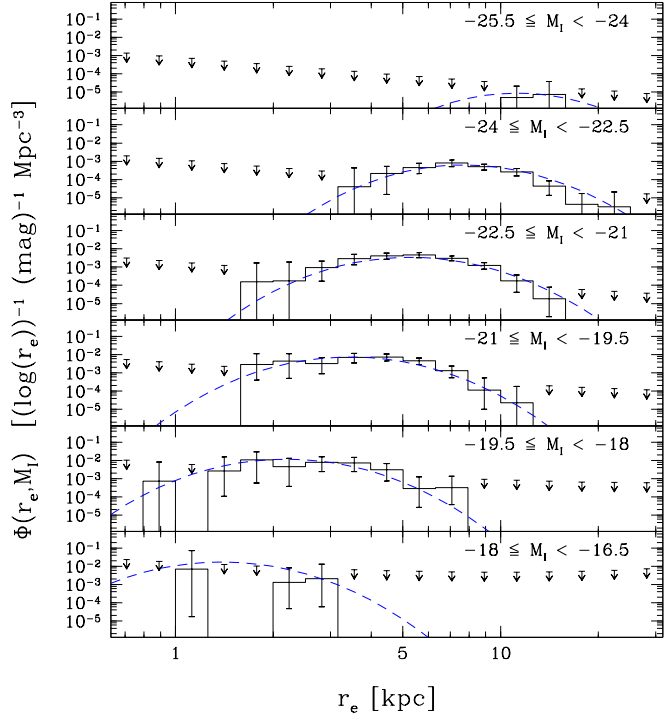}
}
\end{center}
 \caption{
 The space density distribution of effective scalesizes in different
bins of total $I$-band luminosity as marked in the top-right of each
panel.  The histograms represent the data with errorbars showing the
95\% confidence limits due to distance and Poisson errors.  The 95\%
confidence upper limits were calculated using exponential disks
and the survey limits determined from the photometry.  The dashed
line shows the bivariate distribution function described in the text. 
}
 \label{dismagtot_re}
 \end{figure}

These approximations give $\lambda \propto r_{\rm e} L^{-(2/\beta-1)}
\simeq r_{\rm e} L^{-1/3}$.  As $\lambda$ is expected to have a
log-normal behavior, this means that, {\em at a given luminosity, we
expect the distribution of scalesizes to be log-normal, and that the
peak in the \reff\ distribution shifts with $\sim L^{-1/3}$}.  This is
exactly the behavior that is shown in Fig.\,\ref{dismagtot_re}, where
the function over-plotted on the data shows the log-normal behavior at
each luminosity bin, shifting by $L^{-1/3}$ between the luminosity bins
and where the height is determined by the Schechter LF. 

The function plotted is the result of the well known $\chi$-by-eye
fitting method, and the detailed parameters will definitely change when
a full fitting technique has been developed that takes the Poisson
errors on the data points into account.  For reference we list here the
full bivariate function in magnitudes, and the parameter values giving a
good approximation to the data:
 \begin{eqnarray}
 \Phi(r_{\rm e},M)\,d\log r_{\rm e}\,dM&=&\frac{\Phi_0}{\sigma_\lambda \sqrt{2\pi}}
\exp(-\frac{1}{2}[\frac{\log r_{\rm e}/r_{{\rm e}*}-0.4(M-M_*)(2/\beta-1)}{\sigma_\lambda/\ln(10)}]^2)
\nonumber\\[1.3mm]
                    & &10^{-0.4(M-M_*)(\alpha+1)} \,
%\exp(-10^{-0.4(M-M_*)}) \, d\log\frac{r_{\rm e}}{r_{{\rm e}*}}\,dM, \nonumber 
\exp(-10^{-0.4(M-M_*)}) \, d\log r_{\rm e}\,dM, \nonumber 
 \end{eqnarray}
 with the first line representing the log-normal scalesize distribution
and the second line the Schechter LF in magnitudes ($M$).  The
$\chi$-by-eye parameters are:\\[1.5mm]
 \begin{tabular}{lll}
$\Phi_0=0.002 {\rm\, Mpc}^{-3}$ & $\alpha = -1.25$ & $\beta = 3.0$ \ (slope
Tully-Fisher\ relation)\\[1.5mm]
$M_* = -22.3\,I$-mag& $r_{{\rm e}*} = 6.7\,{\rm kpc}$ & $\sigma_\lambda=0.3$
 \end{tabular}\\[1.5mm] 
 The width of the spin parameter distribution ($\sigma_\lambda$) we need
is less than what is typically found in N-body simulations.
%A similar
%conclusion following a different analysis method was reached by Mo et
%al.~(1998).% ***Explanations!?***

%Our parameterization can be transformed in any combination of luminosity,
%SB and scale size, using the relation $M = \mu_0 -
%5\log(h)-2.5\log(2\pi)$ for perfect exponential disks, as long as one
%takes the proper derivatives into account (see also Dalcanton et
%al.~1997).  Figures\,\ref{**} and \ref{**} show therefore exactly the
%same distribution, just in a different projection.

%---------------------------------------------------------------------------
\section{Discussion \& conclusions}

 \begin{figure}
\begin{center}
\mbox{
 \epsfxsize=9cm
 \epsfbox[30 185 550 530]{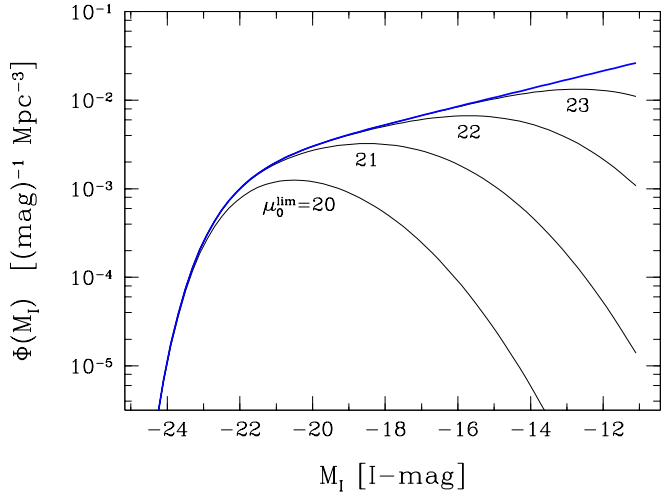}
}
\end{center}
 \caption{
 Luminosity functions computed by integrating the bivariate distribution
function down to the indicated $I$-band central surface
brightness limits.  The uppermost line shows the integrated total LF.
}
 \label{phifuncLF}
 \end{figure}

Using our parameterization we can now estimate the effects of SB limits
on determinations of LFs and on deep galaxy counts, especially in the
context of (1+$z$)$^4$ redshift dimming.  If we were to determine the LF
from a galaxy sample with low SB galaxies cut out, we would
underestimate the faint end of the LF, as most low SB galaxies are also
low luminosity systems.  SB cuts are relevant for redshift surveys
selected from shallow (photographic plate) material (resulting in
implicit cuts) and for most fiber based redshift surveys which often
have explicit SB cuts (e.g.\ the Las Campanas and Sloan
Surveys).  How SB cuts can effect LF determinations is
shown in Fig.\,\ref{phifuncLF}, where we have integrated our
parameterization down to the indicated $I$-band {\em central}\, SB. 

Figure\,\ref{phifuncLF} suggests that the SB cuts typically present in
local surveys do not dramatically effect LF determinations (using for
a typical spiral \hbox{$B$--$I$$\simeq$1.7}), especially taking into account
that most spirals have some central light enhancement due to the bulge,
making detections easier.  The situation changes however when we move to
higher redshifts and have to take (1+$z$)$^4$ cosmological redshift
dimming into account.  At $z$=1 our SB limit has already shifted 3
magnitudes up, and 6 magnitudes by the time we reach $z$=3.  This means
that even for the deepest image available at the moment --the Hubble
Deep Field-- the SB cut at $z$=3 (the $U$-band dropouts) runs at about
21\,$I$-mag\,arcsec$^{-2}$ (using a K-correction of an unevolved Sb
galaxy).  This limit makes a considerable fraction of galaxies in
Fig.\,\ref{bidisremuave} undetectable, if we put this local galaxy
population unevolved at $z$=3. 

Tully \& Verheijen (1997, these proceedings) have argued that the
central SB of galaxies shows a bimodal distribution, in particular when
looking at $K$-band data.  We do not see such bimodality, independent
whether we use their proposed bimodal dust extinction correction, we use
only the 200 most face-on galaxies with the smallest extinction
correction, we use bulge/disk decomposed parameters or effective
parameters.  In the many ways we have looked at the MFB data set, we
have never seen any bimodality in the SB distributions.  Whether the
bimodal effect is the result of the special Ursa Major cluster
environent that was studied
 %(even though a
%fair fraction of the MFB galaxies must lie in the outer
%parts of clusters) 
 or an unlucky case of low number statistics remains
to be seen. 

The simple parametrization presented in this paper gives an accurate
representation of the observed bivariate distributions, independently of
whether one believes in hierarchical galaxy formation models or in
CDM-like universes.  A detailed analysis of galaxy formation in CDM-like
universes paying attention to bivariate space density distributions will
appear in Lacey et al.~(1999).

%---------------------------------------------------------------------------
\acknowledgments

Support for R.S.\ de Jong was provided by NASA through Hubble Fellowship
grant \#HF-01106.01-98A from the Space Telescope Science Institute,
which is operated by the Association of Universities for Research in
Astronomy, Inc., under NASA contract NAS5-26555.
%We thank Vince Ford for making the luminosity profiles of the Mathewson
%et al.\ data set available in machine readable format.

%---------------------------------------------------------------------------


\begin{references}
\reference Byun, Y.-I. 1992, PhD. Thesis, The Australia National University
\reference Dalcanton J. J., Spergel, D. N. \& Summers, F. J. 1997, \apj,
482, 676
\reference de Jong, R. S. 1996, \aap, 313, 45
\reference Disney, M. J. 1976, Nature, 263, 573
\reference Disney, M. J. \& Phillipps, S. 1983, \mnras, 205, 1253
\reference Ellis, R. S. 1997, ARA\&A, 35, 389
\reference Fall, S. M. \& Efstathiou, G. 1980, \mnras, 193, 189
\reference Felten, J. E. 1976, \apj, 207, 700
\reference Freeman, K. C. 1970, \apj, 160, 811
\reference Hudson, M. J. \& Lynden-Bell, D. 1991, \mnras, 252, 219
\reference Impey, C. \& Bothun, G. 1997, ARA\&A, 35, 267
\reference Lacey, C., Cole, S., Baugh, C. \& Frenk, C. S. 1999, in preparation
\reference Mathewson, D. S. \& Ford, V. L. 1996, \apjs, 107, 97
\reference Mathewson, D. S., Ford, V. L. \& Buchorn M. 1992, \apjs, 81, 413
\reference McGaugh, S.S., Bothun, G. D. \& Schombert, J. M. 1995, \aj, 110, 573
\reference Mo, H. J., Mao, S. \& White, S. D. M.  1998, \mnras, 295, 319
\reference Peebles, P. J. E. 1969, \apj, 155, 393
\reference Press, W. H., Schechter, P. 1974, \apj, 187, 425
\reference Schechter, P. 1976, \apj, 203, 297
\reference Schlegel, D. J., Finkbeiner, D. P. \& Davis, M. 1998, \apj, 500, 525
\reference Sodr\'e, L. \& Lahav, O. 1993, \mnras, 260, 285
%\reference Tully, R. B. \& Fisher J. R. 1977 \aap, 54, 661
\reference Tully, R. B. \& Verheijen, M. A. W. 1997, \apj, 484, 145
\reference van den Bosch, F. C. 1998, submitted to \apj, astro-ph/9805113
\reference van der Kruit, P. C. 1987, \aap, 173, 59
\reference Warren, M.S., Quinn, P. J., Salmon, J. K. \& Zurek, W. H.
1992, \apj, 399, 405
\reference Willick, J. A. et al. 1997, \apjs, 109, 333
\end{references}
\end{document}